\documentclass[12pt]{article}
\usepackage[margin=0.95 in]{geometry}
\usepackage{amsmath} 
\usepackage{amssymb,amsfonts}
\usepackage[all]{xy}
\usepackage{graphicx}

\usepackage{amsmath}
\usepackage{amssymb}
\usepackage{float}
\usepackage{array}
\usepackage{tikz}
\usepackage{mathtools}
\usepackage{mathrsfs} 
\usepackage[hidelinks]{hyperref}
\usepackage{cite}
\numberwithin{equation}{section}
\usepackage[utf8x]{inputenc}

\newcommand{\be}{\begin{equation}}
\newcommand{\ee}{\end{equation}}
\def\bea{\begin{eqnarray}}
\def\eea{\end{eqnarray}}

\numberwithin{equation}{section}
\numberwithin{table}{section}

\begin{document}

\hypersetup{pageanchor=false}
\begin{titlepage}
\vbox{
    \halign{#\hfil         \cr
           } 
      }  
\vspace*{15mm}
\begin{center}
{\Large \bf 
Topological Two-Dimensional Gravity}

\vspace*{3mm}

{\Large \bf on Surfaces with Boundary
}

\vspace*{15mm}

{\large  Jan Troost$^b$ }
\vspace*{8mm}

$^b$ Laboratoire de Physique de l'\'Ecole Normale Sup\'erieure \\ 
 \hskip -.05cm
 CNRS, ENS, Universit\'e PSL,  Sorbonne Universit\'e, \\
 Universit\'e de Paris 
 \hskip -.05cm F-75005 Paris, France	 
\vskip 0.8cm
	{\small
		E-mail:
		\texttt{ jan.troost@ens.fr}
	}
\vspace*{0.8cm}
\end{center}

\begin{abstract}
We solve two-dimensional gravity on surfaces with boundary in terms of contact interactions and surface degenerations. The known solution of the bulk theory in terms of a contact algebra is generalized to include boundaries and an enlarged set of boundary operators. The latter  allow for a linearization of the Virasoro constraints in terms of an extended integrable KdV hierarchy.

\end{abstract}

\end{titlepage}
\hypersetup{pageanchor=true}

\tableofcontents

\section{Introduction}
Two-dimensional gravity on closed Riemann surfaces was solved in terms of matrix models \cite{Brezin:1990rb,Douglas:1989ve,Gross:1989vs}, conformal field theory \cite{Knizhnik:1988ak,David:1988hj,Distler:1988jt} and intersection theory \cite{Witten:1989ig,Kontsevich:1992ti}. While aspects of gravity on Riemann surfaces with boundary were partially understood in terms of matrix models early on \cite{Dalley:1992br,Johnson:1993vk}, a rigorous theory of topological gravity on Riemann surfaces with boundary was only recently established \cite{Pandharipande:2014qya}.  Since then, various perspectives on these theories have been developed \cite{Buryak:2014dta,Alexandrov:2014gfa,Buryak:2015eza,Dijkgraaf:2018vnm,Aleshkin:2018klz,Alexandrov:2019wfk}. The main approaches are through geometry and matrix models. The points of view provided by these methods on the resulting integrable KdV hierarchy are qualitatively distinct and usefully complementary.

Two-dimensional gravity on closed Riemann surfaces was also understood in  a conformal field theory approach  closely related to string theory \cite{Verlinde:1990ku}. The theory was solved in terms of Virasoro recursion relations. These relations were derived from a contact algebra for vertex operators that carries all the topological information provided by the surface as well as the bundles on the moduli space of surfaces \cite{Witten:1989ig}. 

Our goal in this paper is to extend the contact algebra approach \cite{Verlinde:1990ku} to topological gravity on Riemann surfaces with boundary. To that end, we study the contact algebra for operators in the presence of boundaries as well as how the bulk algebra is represented on an extended set of boundary vertex operators.  Through representation theory and consistency conditions, we fix all constants in the extended open Virasoro algebra, and manage to derive the Virasoro recursion relation for the open and closed partition functions. Given a few initial correlators, this allows to solve the theory. 

The paper is structured as follows. 
In section \ref{OpenTopologicalGravity} we review salient features of topological gravity on Riemann surfaces with boundary \cite{Pandharipande:2014qya}. The extended representation of the bulk vertex operator contact algebra on the 
boundary vertex operators is constructed in section \ref{ContactAlgebraRepresentation}  using  consistency arguments. In section \ref{Virasoro} the constraints are translated into a differential Virasoro algebra that acts on the generating function of topological correlators. At that point, we make contact with the extended open string partition function \cite{Buryak:2014dta} which is sufficient to prove that the solution to the Virasoro constraints indeed coincides with the known solution of open topological gravity. We conclude in section \ref{Conclusions} with a summary and suggestions for  future research.

\section{Open Topological Gravity}
\label{OpenTopologicalGravity}
In this section, we recall features of the solution of open and closed topological gravity, respectively on Riemann surfaces with \cite{Pandharipande:2014qya} or without boundary \cite{Witten:1989ig}. For open topological gravity, we indicate a few features of the rigorous geometric solution \cite{Pandharipande:2014qya}. For topological gravity on Riemann surfaces (without boundary), we also briefly recall aspects of the solution in terms of a conformal field theory \cite{Verlinde:1990ku} with contact interactions. We then start out on the path to generalize that solution to Riemann surfaces with boundary.\footnote{See also \cite{Gaiotto:2003yb} for an interesting alternative.}

\subsubsection*{Riemann Surfaces and  Carriers of Curvature}
Topological gravity on Riemann surfaces (without boundary) \cite{Witten:1989ig} satisfies the ghost number conservation equation -- or the dimension constraint on the integral over the moduli space of surfaces --:
\begin{equation}
3g-3+n^c = \sum_{i=1}^{n^c} n_i^c \, .
\label{ClosedGhostNumber}
\end{equation}
The genus of the Riemann surface is $g$.
The number of bulk vertex operator insertions is $n^c$ and $n_i^c$ are the labels of the bulk vertex operators referring to the power of the tangent line bundle at a point \cite{Witten:1989ig}.
A central idea in \cite{Verlinde:1990ku} was to graft the curvature associated to the Riemann surface itself onto the bulk vertex operators such that all topological properties of the theory are captured by local operators -- this in turn allows for the solution of the theory in terms of contact interactions. When we associate a curvature $2 (n_i^c-1)/3$ to each bulk vertex operator $\tau_{n_i^c}$  of power $n_i^c$, then ghost number conservation implies that:
\begin{equation}
\text{The Integrated Curvature} = 2g-2 = \sum_{i=1}^{n^c} \frac{2}{3} (n_i^c-1) \, ,
\label{BulkCurvature}
\end{equation}
namely that the curvature of the surface is faithfully represented. The puncture operator $\tau_0$ has the smallest curvature contribution equal to $-2/3$, while the dilaton operator $\tau_1$ carries no curvature at all. All other operators carry positive curvature (in this convention).
\subsubsection*{Riemann Surfaces with Boundary}
The integration over the moduli space of Riemann surfaces with boundaries and with boundary and bulk insertions leads to the dimensionality constraint valid for non-zero open correlation functions \cite{Pandharipande:2014qya}:
\begin{equation}
3g'-3+n^o+2n^c = 2 \sum_{i=1}^{n^c} n_i^c \, .
\end{equation}
 The doubled genus $g'$ is the genus of the Riemann surface that is obtained by gluing a given Riemann surface with at least one boundary to its reflection. We therefore have the relation $g'=2g +b-1 $ where $b$ is the number of boundaries of the original surface and $g$ its genus. The number of boundary operator insertions $\sigma$  is $n^o$ \cite{Pandharipande:2014qya}. In terms of the ordinary genus $g$ and number of boundaries $b$, we have:
\begin{equation}
6g-6+3b+2n^c+n^o= 2 \sum_{i=1}^{n^c} n_i^c \, ,
\label{OpenGhostNumber}
\end{equation}
in which we recognize the constraint (\ref{ClosedGhostNumber}) as the special case without boundaries.

Our first step in generalizing the solution of the closed theory in terms of contact interactions \cite{Verlinde:1990ku} is to appropriately distribute curvature in the presence of boundaries and boundary insertions. We continue to assign curvature to the bulk insertions as before \cite{Verlinde:1990ku} -- see equation (\ref{BulkCurvature}). For simplicity, we momentarily imagine a single boundary, with a non-zero number $n^o$ of boundary insertions $\sigma$. The ghost number conservation equation (\ref{OpenGhostNumber}) then suggests that we should assign curvature $-1/3$ to each basic boundary insertion $\sigma$, in such a manner that we find  the equation:
\begin{equation}
\text{Boundary Curvature} = 1= -\frac{n^o}{3} \, , 
\end{equation}
in accord with our assignment for bulk curvature as well as the ghost number conservation equation (\ref{OpenGhostNumber}). The relative factor of a half compared to the basic bulk (puncture) operator $\tau_0$ is due to the fact that the boundary operator increases the dimension of the moduli space by real dimension one (compared to a bulk operator which increases the real dimension by two). 
This reasoning can be generalized to the case of multiple boundaries with insertions. It is sufficient to introduce an extra label corresponding to each boundary (with its associated boundary insertions). 
We conclude that the boundary operator $\sigma$ carries curvature $-1/3$. 

\subsubsection*{Higher Powers}
\label{PreviewRho}
To prepare for reasonings to come, it may be useful to interject a thought experiment at this point. 
Note that the closed string vertex operator $\tau_n$ can be thought of as a power of the vertex operator $\tau_1$ in an approximate sense. The curvature it carries is then interpreted as the curvature $n \times 2/3$ from which we subtract $2/3$. The curvature remains bounded from below such that the vertex operators do not cut out such a large part of the surface for it to disappear entirely.\footnote{This is dictated by geometry or can be interpreted as a Seiberg bound \cite{Dijkgraaf:1990qw}.}
Similarly, if we were to attempt to define an arbitrary power of the boundary operator $\sigma$ to which we attached curvature $-1/3$, the operator would not have well-defined correlation functions.  A manner to remedy this obstruction is to add explicit powers of the string coupling $u$ to the operator: $\rho_n = u^{n-1} \sigma^n$. Now, the powers of the string coupling are counted by the genus $g$ and the number of boundaries $b$ on the one hand, and the explicit powers of $u$ on the other hand. Suppose we study a correlation function of operators $\rho_{n_j^o}$ and $\tau_{n_i^c}$. It  satisfies the equation:
\begin{equation}
2g-2+b+\sum_{j=1}^{n^o} (n_j^o-1) =  \sum_{i=1}^{n_c} \frac{2(n_i^c-1)}{3} + \sum_{j=1}^{n^o} (\frac{2n_j^o}{3}-1) \, .
\label{OpenGhostNumberRewritten}
\end{equation}
This is still the ghost conservation equation (\ref{OpenGhostNumber}) but rewritten in such a way as to make the explicit string coupling contributions visible on the left hand side. We made use of the fact that the coupling $u$ corresponds to the vacuum expectation value of the exponential of the dilaton operator $\tau_1$ which couples to curvature.
The operator $\rho_n$  still caries ghost number $n$, but it also carries curvature $2n/3-1$, as we made manifest in our manner of writing equation (\ref{OpenGhostNumberRewritten}).\footnote{For  $n \ge 1$ the boundary operator now has sufficient  curvature to have well-defined correlation functions.}
 While the boundary operators  that we will soon encounter are more intricate still, they share features with the operators $\rho_n$.

\section{The Virasoro Algebra Representations}
\label{ContactAlgebraRepresentation}
In this subsection, we briefly remind the reader of an intuitive manner to solve topological gravity on closed Riemann surfaces using contact terms \cite{Verlinde:1990ku}. We extend the approach to include boundaries and boundary vertex operators which can be viewed as representing the contact algebra. This section heavily relies on background provided in \cite{Verlinde:1990ku} to which we do refer for more details.

\subsection{The Bulk Representation of the Virasoro Algebra}
The method of \cite{Verlinde:1990ku} to solve topological gravity on closed Riemann surfaces is to represent all the topological data in terms of local operators in a conformal field theory. As an example, we already saw that the curvature (which codes the genus) was assigned to  local bulk operator insertions.  Intersection numbers are then represented as integrals over the moduli space of the Riemann surface of conformal field theory correlators.\footnote{This is heavily reminiscent of string theory (see e.g. \cite{Polchinski:1998rq}) and we  allow string theory nomenclature to creep into our language.} We denote the curvature carrying bulk local operator insertions $\tau_n$. Due to the topological nature of the theory, the contact interactions between the local operators suffice to compute the intersection numbers on the moduli space of Riemann surfaces. 

The method of \cite{Verlinde:1990ku} to solve topological gravity uses the fact that the algebra of integrated vertex operators is represented on localized bulk vertex operators (or states) in the form \cite{Verlinde:1990ku}:
\begin{equation}
\int_{D_\epsilon} \tau_m | \tau_n \rangle = A^n_m | \tau_{n+m-1} \rangle
\, ,
\end{equation}
where the localized vertex operator $\tau_n$ is assumed to lie in the disk $D_\epsilon$ over which the vertex operator $\tau_m$ is integrated. 
The representation arises from the contact term between the operators $\tau_m$ and $\tau_n$. 
When we wish to compute the algebra of consecutive actions of the locally integrated bulk vertex operators in the representation, we need to take into account that the first integrated operator may enter into contact with the second integrated operator. To keep track of this term, it is useful to define a measure of the non-commutativity of the operation of localizing the vertex operator, and integrating over it \cite{Verlinde:1990ku}:
\begin{equation}
\int_{D_\epsilon} \tau_m | \tau_n \rangle -\int_{D_\epsilon} \tau_n | \tau_m \rangle = C_{nm} | \tau_{n+m-1} \rangle \, .
\end{equation}
 Then, when we consider the action of two integrated vertex operators on a localized operator,
we find a consistency condition between the representation coefficients $A$ and the measure of non-commutativity $C$ \cite{Verlinde:1990ku}:
\begin{eqnarray}
A_n^{m+k-1} A_m^k -A_m^{n+k-1} A_n^k + C_{nm} A_{m+n-1}^k &=& 0 \, . \label{VirasoroRepresentation}
\end{eqnarray}
The coefficient $A^n_m$ is calculated in  \cite{Verlinde:1990ku} and it equals the curvature of the insertion plus one:
\begin{equation}
A^n_m =  \frac{2}{3} (n-1)+1\, , 
\label{CurvaturePlusOne}
\end{equation}
and we retain that we have the contact contribution
\begin{equation}
\label{BulkContactAlgebraRepresentation}
\int_{D_\epsilon} \tau_m | \tau_n \rangle =
\frac{2n+1}{3} | \tau_{n+m-1}  \rangle \, .
\end{equation}
In turn this implies that the measure of non-commutativity $C$ is proportional to the difference in the curvature of the insertions:
\begin{equation}
C_{mn} = \frac{2}{3} (m-n) \, .
\end{equation}
Note that when we identify the coefficients $A_n$ of the representation on the bulk vertex operator space with an operator $L_{n-1}$, then the commutation relation (\ref{VirasoroRepresentation}) shows that we have a  representation of the  Virasoro algebra:
\begin{equation}
[L_n, L_m] = \frac{2}{3} (m-n) L_{m+n} \, .
\end{equation}
Thus, the contact algebra is a Virasoro algebra, represented on the space of bulk operator insertions. This is an essential tool in the solution to the bulk topological gravity theory \cite{Verlinde:1990ku}, and we wish to extend it to Riemann surfaces with boundary. 

\subsection{The Extended Virasoro Representation}
In the presence of a boundary, we first address the question what happens when a bulk vertex operator is integrated over a small ring $R_\epsilon$ near an empty boundary. We propose that the integrated vertex operator in that case generates an operator  on the boundary:
\begin{equation}
\int_{R_\epsilon} \tau_{n} | \,  \rangle^b = u \, c(n) | \sigma_{n-1}^b \rangle \, .
\label{EmptyBoundary}
\end{equation}
We have introduced operators $\sigma_n^b$ that live on a boundary of the Riemann surface. We have stripped off one factor of the string coupling constant $u$ on the right hand side -- we think of the bulk vertex operators as carrying one power of the coupling constant more than the boundary operators.\footnote{This is standard in string theory. Alternatively, it can be viewed as a consequence of the relative contribution of bulk and boundary vertex operators to the dimension of moduli space.} We have allowed for a representation coefficient $c(n)$ that is undetermined for now.  The curvature of the operator
 $\sigma^b_{n}$ equals the  curvature of the bulk vertex operator minus one, to compensate for the string coupling constant prefactor. Therefore, the curvature of the operator $\sigma_{n-1}^b$ equals $2(n-1)/3-1$. We allow for operators with $n \ge 2$ and set other terms to zero. 
 
 Thus, we have introduced a new space parameterized by the operators $\sigma_n^b$. Our next step is to assume that the integrated bulk vertex operators also act on this space and  provide a new representation of the Virasoro algebra. 
 We need to make sure that the resulting operator carries the sum of the curvatures of the operators on the left hand side, and we propose that the contact algebra coefficient is again fixed to equal the curvature of the operator plus one -- see equation (\ref{CurvaturePlusOne}). We thus find:
\begin{align}
\label{BoundaryContactAlgebraRepresentation}
\int_{R_\epsilon} \tau_m | \sigma_{n}^b \rangle
&= 
\frac{2n}{3} | \sigma_{m+n-1}^b \rangle \, .
\end{align}
This natural proposal partially fixes the normalization of the boundary vertex operators.  We still need to check whether the  integrated  vertex operators  satisfy the Virasoro algebra. The action (\ref{BoundaryContactAlgebraRepresentation}) is indeed a representation of the Virasoro algebra, as before. For the action (\ref{EmptyBoundary}) to also enter into a representation of the Virasoro algebra,   the coefficient $c(n)$ needs to be a linear function of $n$. Finally, we use a choice of overall normalization of the boundary vertex operators to set $c(n)= \frac{n+a}{3}$ where $a$ is a constant to be
\label{ConstantA}
determined. We will later argue that consistency requires $a=0$ and we therefore find the action on an empty boundary:
\begin{equation}
\label{BoundaryContactAlgebra}
\int_{R_\epsilon} \tau_{n} | \,  \rangle^b = u \, \frac{n}{3} | \sigma_{n-1}^b \rangle \, .
\end{equation}
In summary, we have extended the space of boundary operators considerably, and we have represented the Virasoro contact algebra on that space. 

\subsection{The  Recursion Relation}
For topological gravity on closed Riemann surfaces, the representation of the contact algebra was leveraged into a recursion relation for the topological correlators \cite{Verlinde:1990ku}. The integral over bulk vertex operators was split into an integral over small disks where other operators reside, neighbourhoods of nodes and uneventful regions. The fact that integrals of bulk operators over the whole Riemann surface  should commute, combined with the contact algebra, gave rise to consistency conditions on the contributions of nodes which in turn provided a recursion relation for correlators. Our claim is that the same reasoning applies to the integrated bulk vertex operators on Riemann surfaces with boundary. We again need to take into account the possible development of nodes on the Riemann surface, as well as possible generalized contact terms with the boundary, which we described previously. 

\begin{figure}
\begin{center}
\includegraphics[width=400pt]{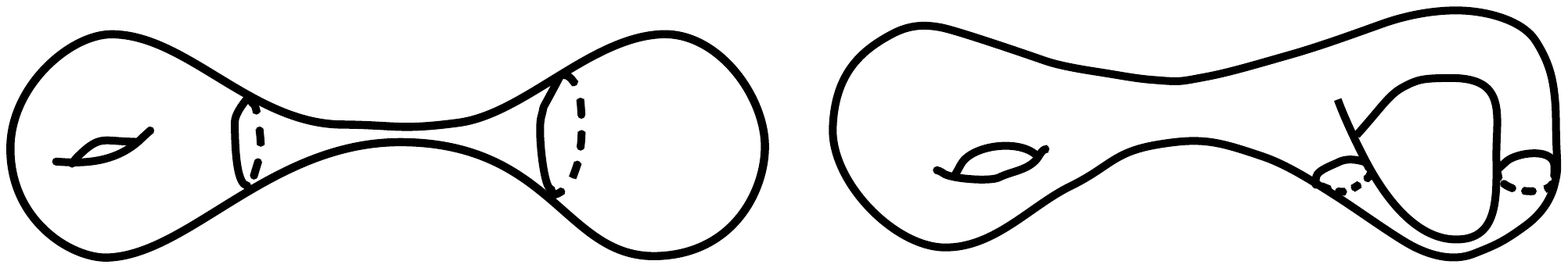}
\caption{Two degenerations of Riemann surfaces are depicted. The left figure represents a surface splitting into two surfaces. The sum of genera is conserved. The right figure shows a genus two Riemann surface that turns into a genus one Riemann surface, lowering the genus by one.}
\label{ClosedTransition}
\end{center}
\end{figure}

To ease into the generalized recursion relation, let us recall the closed recursion relation first \cite{Verlinde:1990ku}\footnote{We normalize the bulk correlators as $\langle \tau_0 \tau_0 \tau_0 \rangle^c=1$ and $\langle \tau_1 \rangle^c=1/24$. We often set the string coupling $u$ to one.}:
\begin{eqnarray}
\langle \tau_{n+1} \prod_{i \in C} \tau_{n_i} \rangle^c
&=& \sum_j \frac{2n_j+1}{3} \langle \tau_{n+n_j} \prod_{i \neq j} \tau_{n_i} \rangle^c
\label{ClosedRecursion}
 \\
&& + \frac{u^2}{18} (\sum_{k=0}^{n-1} ( \langle \tau_k \tau_{n-k-1} \prod_{i \in C} \tau_{n_i} \rangle^c
+ \sum_{C=C_1 \cup C_2} \langle \tau_{k} \prod_{i \in C_1} \tau_{n_i} \rangle^c \langle \tau_{k-i-1} \prod_{j \in C_2} \tau_{n_j} \rangle^c) \nonumber \, .
\end{eqnarray}
The set $C$ is a set of bulk operator insertions. 
 The first term on the right hand side arises from the bulk contact algebra representation (\ref{BulkContactAlgebraRepresentation}) while the second line has its origins in the fact that a Riemann surface can develop nodes which give rise to a Riemann surface of one genus less, or which splits the Riemann surface into two closed Riemann surfaces. See Figure \ref{ClosedTransition} and reference \cite{Verlinde:1990ku}.
 
 The generalization to the case of extended open correlators is:
 \begin{eqnarray}
& & \langle \tau_{n+1} \prod_{i \in C} \tau_{n_i}
\prod_{l \in O}
\sigma^b_{n_l} \rangle^{o,ext}  \nonumber \\
&=&  \sum_j \frac{2n_j+1}{3} \langle \tau_{n+n_j} \prod_{i \neq j} \tau_{n_i} \prod_l \sigma_{n_l} \rangle^{o,ext}
+ \sum_j \frac{2n_j}{3} \langle \prod_{i } \tau_{n_i} \sigma_{n+n_j}  \prod_{l \neq j} \sigma_{n_l} \rangle^{o,ext}  \nonumber 
 \\
& &  + u \frac{n+1}{3} \langle \sigma_n^b \prod_{i \in C} \tau_{n_i}
\prod_{l \in O}
\sigma^b_{n_l} \rangle^{o,ext}  
\label{RecursionRelation}
\\
& & + \frac{u^2}{18} (\sum_{k=0}^{n-1} ( \langle \tau_k \tau_{n-k-1} \prod_{i,j \in CO} \tau_{n_k} \rangle^{o,ext}
\nonumber \\
& & 
+ \sum_{(e,f)} \sum_{CO=CO_1 \cup CO_2} \langle \tau_{k} \prod_{i,j \in CO_1} \tau_{n_i} \sigma_{n_j}^b \rangle^e \langle \tau_{k-i-1} \prod_{l,m \in CO_2} \tau_{n_l} \sigma_{n_m}^{b} \rangle^f) \nonumber  \, .
\end{eqnarray}
The first line corresponds to the fact that we are considering an integrated bulk operator $\tau_{n+1}$. It gives rise to the contact terms in the second line from the bulk contact term (\ref{BulkContactAlgebraRepresentation}) and the boundary contact term (\ref{BoundaryContactAlgebraRepresentation}). The third line arises from the naked boundary term (\ref{BoundaryContactAlgebra}). The fourth line arises from pinching off a handle. The fifth line requires explanation. We sum over the sectors $(e,f)$ which can be either (open,closed), (closed,open) or (open,open).\footnote{We exclude the case with no boundaries from our definition of extended open correlators. See equation (\ref{ClosedRecursion}) for the purely closed correlators.} The first two arise when we split the surface into a closed Riemann surface and a Riemann surface with boundary.\footnote{We effectively obtain a factor of two from these first two sectors.} In that case, the open string sector will contain all the boundary insertions, necessarily.  The third value, (open,open) arises when a node splits the Riemann surface into two Riemann surfaces with boundary. The set $CO$ indicates the set of all bulk and boundary insertions, and we sum over their possible distributions $CO_1$ and $CO_2$ on the two disjoint surfaces.\footnote{ If one labels boundaries, and their associated boundary operators, a finer combinatorics and summation is necessary.}

Note that the second line in the right hand side contains a correlator that is of one order less in the string coupling constant, and the third line a correlator that is down by two orders in the string coupling constant $u$.

\subsection{The Generalized Vertex Operators}

To make further progress, we must discuss the nature of the extended set of boundary vertex operators $\sigma_n^b$ in more detail. We recall that in the geometric open topological theory \cite{Pandharipande:2014qya}, we found a single boundary vertex operator $\sigma$ of curvature $-1/3$ in section \ref{OpenTopologicalGravity}. This matches the curvature of $\sigma_1^b$ and we will indeed identify the two operators: $\sigma=\sigma_1^b$.\footnote{There is a possible normalization factor between these two operators. Our previous choice of overall normalization of the boundary operators makes sure that this identification is spot on in standard conventions.} The curvature of the general operator $\sigma_n^b$ is $2n/3-1$. To make such operators on the boundary, we can use a power of the operator $\sigma$ as well as the string coupling constant (effectively of curvature one). A natural guess is that there is a component $  \rho_n =u^{-1} (u \sigma)^n$ to the boundary vertex operator $\sigma_n^b$ (as previewed in section \ref{PreviewRho}). However, we also need to allow for more drastic processes.

\begin{figure}
\begin{center}
\includegraphics[width=400pt]{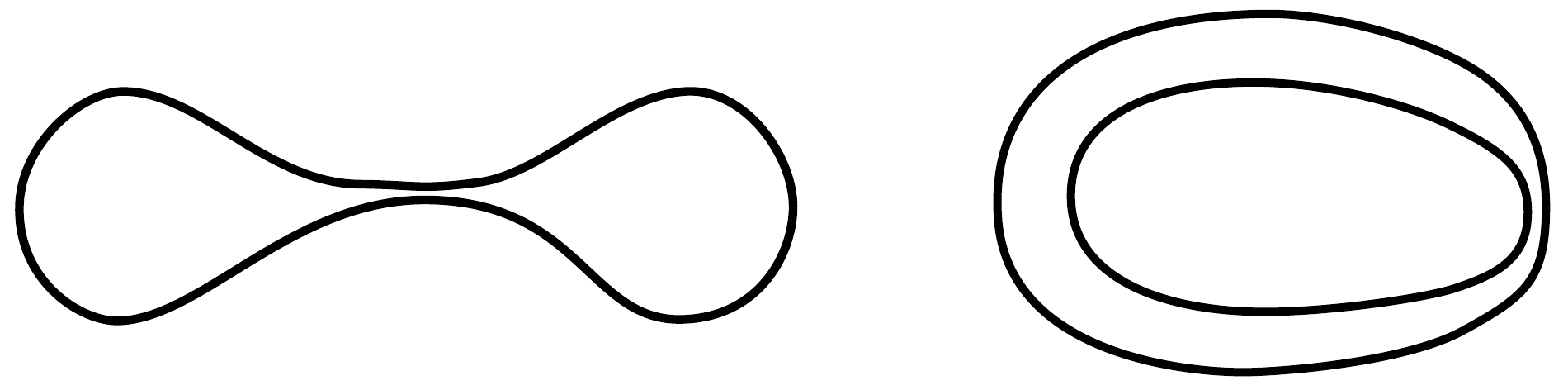}
\caption{Two degenerations of Riemann surfaces with boundary are drawn. The left figure represents a disk splitting into two disks. The right figure shows an annulus that turns into a disk.}
\label{OpenTransition}
\end{center}
\end{figure}

Up to now, a number of complications were implicit in our extended boundary vertex operators. To start with, we concentrate on the simplest extended operator, namely $\sigma_2^b$. It naively corresponds to an insertion of $u \sigma \sigma$. 
However, to understand further possibilities, we need to study the boundary analogue of nodes. A strip (or open string propagator) can be squeezed near the boundary of the moduli space of open Riemann surfaces, in various manners. Either the number of boundaries can decrease as in an annulus to disk transition, or the number of boundaries can increase as in a disk to two disks transition.\footnote{There is a third degeneration process in which the genus drops by one. When one labels boundary components, it will play a role. See e.g. \cite{Zwiebach:1997fe} for a discussion in open/closed string field theory.}  See figure \ref{OpenTransition}. When the integrated bulk vertex operator is close to these singular configurations, it can either give rise to boundary vertex operators that sit on a single boundary or it can give rise to boundary vertex operators that sit on two different boundaries of disconnected surfaces. The  boundary vertex operator $\sigma_2^b$ must capture both these possibilities. Thus, we propose the equation:
\begin{equation}
\langle \dots \sigma_2^b \dots \rangle^{o,ext} = b_1 u \langle \dots \sigma \sigma   \rangle^{o,ext} + b_2 u \langle \dots \sigma  \rangle \langle \sigma \dots \rangle^{o,ext} \, . 
\label{Sigma2Split}
\end{equation}
This equation shows that the generalized vertex operator $\sigma_2^b$ exhibits a non-local characteristic.

We recall that in the case of a node degeneration (see Figure \ref{ClosedTransition}), there was a universality between losing a handle and splitting a surface -- both terms have equal coefficient in the second line of equation (\ref{ClosedRecursion}). 
We propose a similar universality here for the two terms in which the boundary operators remain on the same boundary, or split -- compare Figures \ref{ClosedTransition} and \ref{OpenTransition} --  and set the two constants in the above equation equal, namely $b_1=b=b_2$.  
To determine the overall constant $b$, we calculate an amplitude.

\subsection{Amplitudes}
To understand the content of the recursion relation further, we need initial conditions, which we take from the most basic geometric calculations \cite{Pandharipande:2014qya}. We have that the boundary three-point function is the only non-zero disk amplitude with only boundary $\sigma$ insertions, and normalize it to one:\footnote{This is a disk amplitude. We have set $u=1$ once more.} 
\begin{equation}
\langle \sigma \sigma \sigma \rangle^{o,ext} =1 \, .
\end{equation}
The other initial condition is that the bulk-boundary one-point function on the disk equals:
\begin{equation}
\langle \tau_0 \sigma \rangle^{o,ext} = 1 \, .
\end{equation}
To save on indices, we will drop the upper index on the correlator from now on -- it should be clear from the context which correlator we have in mind.

To understand the structure of the vertex operator $\sigma_{m \ge 2}^b$, we can use the puncture equation, namely, the recursion relation (\ref{RecursionRelation}) for $n=-1$:
\begin{equation}
 \langle \tau_{0} \prod_{i \in C} \tau_{n_i}
\prod_{l \in O}
\sigma^b_{n_l} \rangle
=  \sum_j \frac{2n_j+1}{3} \langle \tau_{n_j-1} \prod_{i \neq j} \tau_{n_i} \prod_l \sigma_{n_l} \rangle
+ \sum_j \frac{2n_j}{3} \langle \prod_{i } \tau_{n_i} \sigma_{n_j-1}  \prod_{l \neq j} \sigma_{n_l} \rangle
   \label{PunctureEquation} \, .
\end{equation}
Let us also be explicit about the  dilaton equation:
 \begin{equation}
\langle \tau_{1} \prod_{i \in C} \tau_{n_i}
\prod_{l \in O}
\sigma^b_{n_l} \rangle  
= \sum_j \frac{2n_j+1}{3} \langle  \prod_{i } \tau_{n_i} \prod_l \sigma_{n_l} \rangle
+ \sum_j \frac{2n_j}{3} \langle \prod_{i } \tau_{n_i}   \prod_{l } \sigma_{n_l} \rangle
  \, .
\end{equation}
We are ready to calculate a first amplitude in two manners, using either the puncture equation, or the factorization equation (\ref{Sigma2Split}):
\begin{eqnarray}
\langle \tau_0 \sigma_2 \sigma \sigma \rangle &=& \frac{4}{3} \langle \sigma \sigma \sigma \rangle 
\nonumber \\
&=& 2 b \langle \tau_0 \sigma \rangle \langle \sigma \sigma \sigma \rangle \, .
\end{eqnarray}
In the first line we used the puncture equation (\ref{PunctureEquation}). In the second line, we used the ansatz (\ref{Sigma2Split}) and allowed for the two possible ways in which the vertex operators can split over two correlators to give a non-vanishing result.\footnote{Ghost number conservation applies to each factor separately.} Note that in the second line a factor of the string coupling constant implicitly cancelled between the two disk amplitudes and the expression for the operator $\sigma_2^b$. 
Using the normalization of the initial conditions, we find:
\begin{equation}
b = \frac{2}{3} \, . 
\end{equation}
This fixes our reading of the extended vertex operator $\sigma_2^b$ once and for all.
For the next extended operator $\sigma_3^b$ we propose a similar universal ansatz consistent with curvature conservation and splitting off a single vertex operator $\sigma$:
\begin{equation}
\langle \dots \sigma_3^b \dots \rangle = c (u \langle \dots \sigma \sigma_2   \rangle +  u \langle \dots \sigma_2  \rangle \langle \sigma \dots \rangle) \, . 
\label{Sigma3Split}
\end{equation}
We can again determine the constant $c$ using either the puncture or the factorization equation to determine one and the same amplitude consistently:
\begin{eqnarray}
\langle \tau_0 \sigma_3 \sigma^4 \rangle &=& 2  \langle \sigma_2 \sigma^4 \rangle = 8 \langle \sigma^3 \rangle \langle \sigma^3 \rangle
\nonumber \\
&=&  c\langle \tau_0 \sigma \rangle \langle \sigma_2 \sigma^4 \rangle + 6 c   \langle \tau_0 \sigma_2 \sigma^2 \rangle \langle \sigma^3 \rangle
\nonumber \\
&=& 4c \langle \tau_0 \sigma \rangle \langle \sigma^3 \rangle \langle \sigma^3 \rangle 
+ 8 c  \langle \sigma^3 \rangle\langle \sigma^3 \rangle \, ,
\end{eqnarray}
and find that again $c=2/3$ -- the constant is fixed once more  in terms of  the bulk-boundary one-point function $\langle \tau_0 \sigma \rangle$. Continuing recursively in this manner, e.g. exploiting the correlation functions $\langle \tau_0 \sigma_n^b \sigma^{2(n-1)} \rangle$,  we find:
\begin{equation}
\label{ReductionExtendedBoundaryOperators}
\langle \dots \sigma_n^b \rangle = u \,  \frac{2 }{3}
(\langle \dots \sigma \sigma_{n-1}^b \rangle + \langle \dots \sigma_{n-1}^b \rangle \langle \sigma \dots \rangle ) \, .
\end{equation}
Thus, we have determined the intricate nature of the extended boundary vertex operators $\sigma_n^b$ and how they recursively code the splitting of boundaries of open Riemann surfaces. 

\subsubsection*{Tying up a loose end: fixing the constant $a$}
We tie up a loose end at the hand of another amplitude.  The amplitude illustrates a splitting of open Riemann surfaces involving two disk one-point functions.
We calculate the amplitude $\langle \tau_3 \tau_0 \sigma \sigma \rangle$ in two manners. We can apply recursion to the operator $\tau_3$, or to the operator $\tau_0$ first. In this calculation, we restore the possible constant $a$ that we introduced in subsection \ref{ConstantA} and use an appropriately modified recursion relation. We  demonstrate that the constant can be determined by consistency. Using the $a$-modified recursion relation, we find:
\begin{eqnarray}
\langle \tau_3 \tau_0 \sigma^2 \rangle &=& \frac{7}{3} \langle \tau_2 \sigma^2 \rangle
\nonumber \\
&=& \frac{1}{3} \langle \tau_2 \sigma^2 \rangle
+ \frac{2}{9} \langle \tau_0 \sigma \rangle \langle \tau_1 \tau_0 \sigma \rangle + \frac{4}{3} \langle \tau_0 \sigma_3 \sigma \rangle + \frac{3+a}{3} \langle \sigma_2 \tau_0 \sigma^2 \rangle \, .
\end{eqnarray}
This implies:
\begin{eqnarray}
\langle \tau_2 \sigma^2 \rangle &=& \frac{1}{9}  \langle \tau_0 \sigma \rangle \langle \tau_0 \sigma \rangle + \frac{4}{3} \langle \sigma_2 \sigma \rangle + \frac{3+a}{3} \frac{2}{3} \langle \sigma^3 \rangle \, .
\end{eqnarray}
We can compute the latter correlator in another manner, using the puncture equation and the modified recursion relation:
\begin{eqnarray}
\langle \tau_2 \sigma^2 \rangle &=& \frac{1}{9} \langle \tau_0 \sigma \rangle \langle \tau_0 \sigma \rangle + \frac{4}{3} \langle \sigma_2 \sigma \rangle+ \frac{2+a}{3} \langle \sigma^3 \rangle \, .
\end{eqnarray}
Using our previous results, we find full consistency if and only if $a=0$. Thus, we tied up the loose end in subsection \ref{ConstantA}.

\section{The Extended Partition Function}
\label{Virasoro}
In this section we introduce the generating function of extended open string correlators and prove that the recursion relations for the correlators imply  Virasoro constraints on the generating function. This allows us to make our results  more rigorous by connecting to the mathematics literature on the integrable structure of the intersection theory on moduli spaces of Riemann surfaces with boundary \cite{Buryak:2014dta}. We conclude the section with a few example amplitudes.

\subsection{The  Generating Function}
 We recall the generating functions of  closed as well as  open topological gravity correlation functions \cite{Pandharipande:2014qya}:
 \begin{eqnarray}
F^c &=& \sum_{g \ge 0, n \ge 1, 2g-2+n>0}
\frac{u^{2g-2}}{n!} \sum_{k_i \ge 0} \langle \tau_{k_1} \dots \tau_{k_n} \rangle_g^c t_{k_1} \dots t_{k_n} 
\nonumber \\
F^{o,geom} &=& \sum_{g',k,l \ge0, 2g'-2+k+2l >0} \sum_{a_i \ge 0} \frac{u^{g'-1}}{k! l!} \langle \tau_{a_1} \dots \tau_{a_l} \sigma^k \rangle_{g'}^{o} s^k \prod_{i=1}^l t_{a_i} \, .
\end{eqnarray}
In view of our enlarged space of boundary vertex operators, 
we also introduce a generating function for extended open topological gravity correlation functions:
\begin{eqnarray}
F^{o,ext} &=&  \sum_{g',k,l \ge 0, 2g'-2+k+2l >0} \sum_{a_i,b_i \ge 0} \frac{u^{g'-1}}{k! l!} \langle \tau_{a_1} \dots \tau_{a_l} \sigma^b_{b_1} \dots \sigma^b_{b_k} \rangle_g^{o,ext} \prod_i t_{a_i} \prod_j s_{b_j} \, .
\end{eqnarray}
\subsubsection*{The Extended Virasoro Constraints}
We define Virasoro generators
\begin{eqnarray}
L_n &=& \sum_{i \ge 0} \frac{2i+1}{2} t_i \partial_{t_{i+n}} - \frac{3}{2} \partial_{t_{n+1}} + \frac{u^2}{12} \sum_{i=0}^{n-1} \partial_{t_i} \partial_{t_{n-i-1}} + \frac{3}{4} \frac{t_0^2}{u^2} \delta_{n,-1} +  \frac{1}{16} \delta_{n,0} \label{ClosedVirasoro}
\\
L_n^{ext} &=& L_n + \sum_{i \ge 0} (i+1) s_{i+1} \partial_{s_{n+i+1}} + u \frac{n+1}{2} \partial_{s_n}
+  \frac{3}{2} \frac{s_1}{u} \delta_{n,-1} + \frac{3}{4} \delta_{n,0} \label{ExtendedVirasoro}
\end{eqnarray}
for $n \ge -1$. These are defined such that
the recursion relation (\ref{ClosedRecursion}) on the closed as well as the recursion relation (\ref{RecursionRelation}) on the extended open correlators leads to the constraints:
\begin{eqnarray}
L_n \exp{F^c} &=& 0
\nonumber \\
L_n^{ext} \exp ( F^c+F^{o,ext} ) &=& 0 \, .
\end{eqnarray}
The extra constants terms in the closed Virasoro algebra (\ref{ClosedVirasoro}) are due to the initialization cases $\langle \tau_0^3 \rangle^c =1 = 24  \, \langle \tau_1 \rangle$ at genus zero and one respectively, while the initial conditions $\langle \sigma^3 \rangle  =1 = \langle \tau_0 \sigma \rangle $ on the disk lead to the extra constants in the extended Virasoro algebra (\ref{ExtendedVirasoro}), which satisfies\footnote{These generators are rescaled by a factor of $2/3$ compared to section \ref{ContactAlgebraRepresentation} in order to reach a standard normalization for the Virasoro algebra.}
\begin{equation}
[ L_m, L_n ] = (m-n) L_{m+n} \, .
\end{equation}
At this stage, we are  able to make contact with rigorous results -- these constraints on an extended partition function of open topological correlators defined through an extended (or unconstrained) integrable KdV hierarchy were found to hold in \cite{Buryak:2014dta}.\footnote{
\label{Normalization}
The translation of variables and normalizations is: $L_n^{there,ext} = (3/2)^n L_n^{ext}$, $t_n^{there} = 3^{-n} (2n+1)!! t_n $ and $s_{n-1}^{there} = (2/3)^{n-1} n! s_n$.}
The relation between the operators $\sigma^b_n$ and $\sigma$ as well as the string coupling constant is neatly captured by a relation between derivatives of the extended partition function:
\begin{equation}
\partial_{s_{n}} = (\frac{2 u }{3})^{n-1}  \partial_{s_1}^n \, .
\end{equation}
This equation was proven from the KdV integrable hierarchy perspective in \cite{Buryak:2014dta}. Using this equation, and setting extended open times $s_{n \ge2}$ to zero, this relation between derivatives imply the higher order  Virasoro constraints on the geometric open topological partition function, where the open Virasoro generators are \cite{Pandharipande:2014qya}:
\begin{equation}
L_n^{o} = L_n + (\frac{2u}{3})^{n} \partial_{s_1}^{n+1} + \frac{n+1}{2} u (\frac{2u}{3})^{n-1} \partial_{s_1}^n + \delta_{n,-1} \frac{3}{2} \frac{s_1}{u} + \delta_{n,0} \frac{3}{4}
\, .
\end{equation}
The Virasoro constraints and the initialization condition are sufficient to determine the full partition function \cite{Pandharipande:2014qya,Buryak:2014dta}.  Through the generating function of extended correlators, we have connected our  arguments with rigorous  results on intersection theory on moduli spaces of Riemann surfaces with boundary \cite{Pandharipande:2014qya,Buryak:2014dta}.

\subsection{A Few More Amplitudes}
\label{Amplitudes}
For illustrative purposes, we  calculate a few more amplitudes. They render the integrable hierarchy structure, the Virasoro constraints and how to solve them more concrete.
\subsubsection{Amplitudes on The Disk}
We have already indicated that on the disk only the third power of the elementary boundary vertex operator $\sigma$ has a non-zero correlation function and equals one, $\langle \sigma^3 \rangle = 1$. The disk bulk-boundary one-point function $\langle \tau_0 \sigma \rangle$ is also one by a choice of normalization. 
 Amplitudes involving extended boundary vertex operators are computed through the reduction formula (\ref{ReductionExtendedBoundaryOperators}). 
A non-trivial example is:
\begin{eqnarray}
\langle \tau_2 \sigma^5 \rangle &=& \frac{10}{3} \langle
\sigma_2^b \sigma^4 \rangle = \frac{40}{3} \, , 
\end{eqnarray}
where we used the recursion relations (\ref{RecursionRelation}) and (\ref{ReductionExtendedBoundaryOperators}) as well as the $6$ choices of factorization. 
After taking into account the different normalization in footnote \ref{Normalization}, this agrees with a  more generic formula in \cite{Pandharipande:2014qya}. 
Another interesting correlation function is $\langle \tau_2 \tau_0 \sigma \sigma \rangle$. It can be computed through the puncture equation (in the first line below) and/or the $L_1$ constraint (in the second line below):
\begin{eqnarray}
\langle \tau_2 \tau_0 \sigma \sigma \sigma \rangle &=& \frac{5}{3} \langle \tau_1 \sigma \sigma \sigma \rangle = \frac{10}{3} \langle \sigma \sigma \sigma \rangle
\nonumber \\
&=& \frac{1}{3} \langle \tau_1 \sigma \sigma \sigma \rangle + 2 \langle \tau_0 \sigma_2 \sigma \sigma \rangle
\nonumber \\
&=& \frac{2}{3} \langle \sigma \sigma \sigma \rangle + 2 \times \frac{8}{3} \langle \sigma \sigma \sigma \rangle = \frac{10}{3} \langle \sigma \sigma \sigma \rangle \, .
\end{eqnarray}
The two ways of computing are in  agreement.

\subsubsection{Higher Order Amplitudes}
Amplitudes that are higher order in the string coupling exhibit qualitatively new phenomena. We illustrate a few. We first compute amplitudes corresponding to cylinder diagrams, with two boundaries and genus zero. 
An interesting amplitude that involves a closed-open factorization due to a node can once again be computed in two manners:
\begin{eqnarray}
\langle \tau_2 \tau_0 \tau_0 \sigma \rangle &=&
\frac{5}{3} \langle \tau_1 \tau_0 \sigma \rangle = \frac{5}{3} \langle \tau_0 \sigma \rangle
\nonumber \\
&=& \frac{2}{3} \langle \tau_1 \tau_0 \sigma \rangle
+ \frac{1}{9} \langle \tau_0^3 \rangle^c \langle \tau_0 \sigma \rangle + \frac{2}{3} \langle \tau_0 \tau_0 \sigma_2 \rangle
\nonumber \\
&=& \frac{2}{3} \langle  \tau_0 \sigma \rangle
+ \frac{1}{9} \langle \tau_0^3 \rangle^c \langle \tau_0 \sigma \rangle + \frac{8}{9} \langle \tau_0 \sigma \rangle \langle \tau_0 \sigma \rangle \, .
\end{eqnarray}
Both ways of computing the correlator lead to the same result, given the normalization of the closed three-point function $\langle \tau_0^3 \rangle^c$ as well as the bulk-boundary one-point function $\langle \tau_0 \sigma \rangle$. 
Finally, we compute an order $O(u^1)$ amplitude. It involves the one-loop closed one-point function $\langle \tau_1 \rangle^c$:
\begin{align}
\langle \tau_3 \sigma \rangle &= \frac{2}{3} \langle \sigma_3 \rangle +  \langle \sigma_2 \sigma \rangle + \frac{1}{9} ( 1+ \langle \tau_1 \rangle ) \langle \tau_0 \sigma \rangle
&=&  ( (\frac{2}{3})^3 + \frac{2}{3}) \langle \sigma^3 \rangle + \frac{1}{9} ( 1+ \langle \tau_1 \rangle ) \langle \tau_0 \sigma \rangle \, .
\end{align}
Needless to say, many more results can be generated, e.g. by computer. We provided a few telling illustrations that provide insight into the foundation of the integrable hierarchy.

\section{Conclusions}
\label{Conclusions}

In the spirit of the solution of the bulk theory \cite{Verlinde:1990ku} and
building on earlier mathematical work \cite{Pandharipande:2014qya,Buryak:2014dta}, we have solved two-dimensional topological gravity on Riemann surfaces with boundary. By making use of an extended set of boundary vertex operators, we rendered the representation of the contact algebra on the boundary linear. Only in a second step the more complicated degeneration of surfaces with boundary is taken into account and the non-linear realization of the (half) Virasoro algebra is found \cite{Buryak:2014dta}. The picture in which the solution of the theory is provided through contact interactions is a welcome intuitive complement to the geometric and matrix model approaches. 

While we have provided a compelling global picture, there are many details that remain to be worked out. It would be good to find the geometric counterpart to the extended set of boundary operators. The link between (the expectation values of) the conformal field theory fields implicit in our analysis \cite{Verlinde:1990ku} and the sections of vector bundles of open topological gravity can be clarified (e.g. by exploiting  references \cite{Dijkgraaf:2018vnm,Dijkgraaf:1990qw}). The analysis of the contact terms in terms of an integration over a degeneration region of the moduli space of open Riemann surfaces would be interesting. It will also be instructive to compare our analysis to  the geometric derivation of the  topological recursion relation  through closed and open factorization \cite{Pandharipande:2014qya}, intuitively reviewed in \cite{Dijkgraaf:2018vnm}.

Another research direction is to exploit the insights developed here and apply them to more general theories.
The generalization to  the extended closed theory \cite{BCT1}) comes to mind, but mostly
to open spin $r$ curves. Geometric \cite{Faber:2006gca}, integrable \cite{Bertola:2014yka,Buryak:2018ypm},  matrix model \cite{Brezin:2012uc,Ashok:2019fxv} and
conformal field theory  
 insights \cite{Muraki:2018rqv} could be  complemented by the perspective developed in this paper.  

The study of these topological theories of gravity is worthwhile in its own right. It occasionally fruitfully interfaces with recent developments. For instance, the KdV integrable hierarchy governing topological gravity also permeates  the  two-dimensional JT-gravity holographic dual of a peculiar (SYK) one-dimensional quantum system -- see e.g. \cite{Okuyama:2019xbv}  and references thereto.  We believe that the further study of these elementary solvable systems, their integrable hierarchy but also their various manifestations in superficially different mathematical structures like topology, matrices and conformal field theory is worthwhile, and may eventually contribute to our understanding of quantum gravity.

\bibliographystyle{JHEP}

\end{document}